\documentclass[12pt,a4paper]{article}

\usepackage{graphicx}

\input epsf

\newcommand{\frat}[2]{\frac{\textstyle #1}{\textstyle #2}}
\newcommand{\vf}[1]{\mbox{\boldmath $#1$}}

\begin{document}

\begin{center}
{\Large \bf Some peculiarities in response to filling up the Fermi
sphere with quarks}\\
 \vspace{0.5cm}
S.V. Molodtsov$^{1,2}$,  G.M. Zinovjev$^{3}$\\
\vspace{0.5cm}
{\small $^1$Joint Institute for Nuclear Research, Dubna, Moscow region,
RUSSIA}\\
\vspace{0.5cm}
{\small $^2$Institute of Theoretical and Experimental Physics, Moscow, RUSSIA}\\
\vspace{0.5cm}
{\small $^3$Bogolyubov Institute for Theoretical Physics,
National Academy of Sciences of Ukraine, Kiev, UKRAINE}
\end{center}
\vspace{0.5cm}

\begin{center}
\begin{tabular}{p{16cm}}
{\small{Considering quarks as the quasi-particles of the model Hamiltonian
with four-fermion interaction we study response to the process of filling up
the Fermi sphere with quarks. }}
\end{tabular}
\end{center}
\vspace{0.5cm}

In the present paper we analyse the process of filling up the
Fermi sphere with quarks
modelling the corresponding Sletter determinant in the direct way.
The quarks are considered as the quasiparticles of the model Hamiltonian with
four-fermion interaction. In Ref. \cite{MZ} the ground state of this model
Hamiltonian was studied in detail and the singularity of mean energy functional
depending on current quark
mass was found out. In the course of this study we need to trace back the
alterations
of respective dressing transformations. It will be seen, from what follows, this
problem is
fairly difficult to treat with well developed machinery of the Green functions
\cite{adg} in which a system
is described by the Hamiltonian with additional term controlling the particle
number $\mu$ $\bar
q \gamma_4 q$ ($\mu$ is the chemical potential of bare quarks). However, the
quark chemical potential conception runs with some uncertainties in interpreting
a baryon chemical potential \cite{bub} and quantitatively produces the
arbitrariness of order $50$ -- $100$ MeV for the phase diagram. Furthermore,
each kind of quarks requires its particular chemical potential. Here we also
consider the impact of filling up the Fermi sphere on the process of
quasi-particle formation. Actually, the approach proposed should be
appended by an analysis of the bound states influence but here we formulate this
problem only doing hope
to get a justification of such a position while completing our consideration.

We start with briefly reminding how the quasiparticle concept appears in this
approach. In Ref. \cite{MZ} the ground state of Hamiltonian with four-fermion
interaction
\begin{equation}
\label{1}
{\cal H}=-\bar q~(i{\vf \gamma}{\vf \nabla}+im)~q-\bar q~t^a\gamma_\mu~q~
\int d{\vf y} ~\bar q'~t^b\gamma_\nu~q'~\langle A^{a}_\mu A^{b}_\nu\rangle({\vf
x}-{\vf y})~,
\end{equation}
taken in the form of a product of two colored currents located in the spatial
points ${\vf x}$ and ${\vf y}$ connected by a formfactor was investigated and
here $q=q({\vf x})$, $\bar q=\bar q({\vf x})$, $q'=q({\vf y})$,
$\bar q'=\bar q({\vf y})$---are the (anti-)quark operators,
\begin{equation}
\label{2}
q_{\alpha i}({\vf x})=\int\frat{d {\vf p}}{(2\pi)^3} \frat{1}{(2|p_4|)^{1/2}}~
\left[~a({\vf p},s,c)~u_{\alpha i}({\vf p},s,c)~ e^{i{\vf p}{\vf x}}+
b^+({\vf p},s,c)~v_{\alpha i}({\vf p},s,c)~ e^{-i{\vf p}{\vf x}}\right]~,
\end{equation}
where $p_4^2=-{\vf p}^2-m^2$, $i$---is the colour index, $\alpha$ is the
spinor index in the coordinate space, $a^+$, $a$ and $b^+$, $b$ are the creation
and annihilation operators of quarks and anti-quarks, $a~|0\rangle=0$,
$b~|0\rangle=0$ and $|0\rangle$ is the vacuum state of free Hamiltonian.
Everywhere that a summation over indices $s$ and $c$ is meant, the index $s$
describes two spin polarization of the quark and index $c$ plays the similar
role for the colour. $t^a=\lambda^a/2$ are the generators of colour gauge group
$SU(N_c)$ and $m$ is the current quark mass. The Hamiltonian density is
considered in the Euclidean space and $\gamma_\mu$ denotes the Hermitian Dirac
matrices,
$\mu,\nu=1,2,3,4$. $\langle A^{a}_\mu A^{b}_\nu\rangle({\vf x}-{\vf y})$ stands
for the formfactor
and its form will be discussed later. The effective Hamiltonian (\ref{1}) arises
as a result of averaging the system of quarks influenced by intensive stochastic
gluon field
$A^a_\mu$, see Ref. \cite{MZ} (for example, by an (anti-)instanton ensemble).
Based on the
form of induced four-fermion interaction the ground state of the system was
searched as the
Bogolyubov trial function composed by the quark--anti-quark pairs with opposite
momenta and with
vacuum quantum numbers \cite{3}, i.e.
\begin{eqnarray}
\label{3}
&&|\sigma\rangle=T~|0\rangle~,\nonumber\\ [-.2cm]
\\ [-.25cm]
&&
T=\Pi_{ p,s}~\exp\{~\varphi~[~a^+({\vf p},s)~b^+(-{\vf p},s)+
a({\vf p},s)~b(-{\vf p},s)~]~\}~.\nonumber
\end{eqnarray}
In this formula and below, in order to simplify the notations, we refer to only
one complicated index which means both polarization indices (spin and colour).
The parameter  $\varphi({\vf p})$ describing the pairing strength is determined
by the minimum of mean energy
\begin{equation}
\label{4}
E=\langle\sigma|H|\sigma\rangle~.
\end{equation}
Introducing the dressing transformation $T$ we define the creation and
annihilation operators of quasiparticles as $A=T~a~T^{-1}$, $B^+=T~b^+T^{-1}$.
Now the quark field operators are presented in the following form
\begin{eqnarray}
\label{5}
&&q({\vf x})=\int\frat{d {\vf p}}{(2\pi)^3} \frat{1}{(2|p_4|)^{1/2}}~
\left[~A({\vf p},s)~U({\vf p},s)~e^{i{\vf p}{\vf x}}+
B^+({\vf p},s)~V({\vf p},s)~ e^{-i{\vf p}{\vf x}}\right]~,\nonumber\\
&&\bar q({\vf x})=\int\frat{d {\vf p}}{(2\pi)^3} \frat{1}{(2|p_4|)^{1/2}}~
\left[~A^+({\vf p},s)~\overline{U}({\vf p},s)~e^{-i{\vf p}{\vf x}}+
B({\vf p},s)~\overline{V}({\vf p},s)~ e^{i{\vf p}{\vf x}}\right]~,\nonumber
\end{eqnarray}
and the spinors $U$ and $V$ are given as
\begin{eqnarray}
\label{6}
&&U({\vf p},s)=\cos(\varphi)~u({\vf p},s)-\sin(\varphi)~v(-{\vf
p},s)~,\nonumber\\ [-.2cm]
\\ [-.25cm]
&&V({\vf p},s)=\sin(\varphi)~u(-{\vf p},s)+\cos(\varphi)~v({\vf p},s)~.\nonumber
\end{eqnarray}
where $\overline{U}({\vf p},s)=U^+({\vf p},s)~\gamma_4$,
$\overline{V}({\vf p},s)=V^+({\vf p},s)~\gamma_4$ are the Dirac conjugated
spinors.

Then the problem of our interest can be formulated in the following way. We need
to construct the state filled in by quasi-particles (the Sletter determinant)
\begin{equation}
\label{7}
|N\rangle=\prod_{|{\mbox{\scriptsize{\vf P}}}|<P_F;S}~A^+({\vf
P};S)~|\sigma\rangle~,
\end{equation}
which possesses the minimal mean energy over state $|N\rangle$. The polarization
indices run here over all permissible values. The momenta and polarizations of
quasi-particles filling in the Fermi sphere are marked by the capital letters.
The free Hamiltonian written in the quasi-particle operators looks like in the
following
\begin{eqnarray}
\label{8}
&&H_0=-\int d {\vf x}~ \bar q({\vf x})~(i {\vf \gamma} {\vf \nabla}+im)~q({\vf
x})=\nonumber\\
&&~~~~~=\int  \frat{d {\vf p}}{(2\pi)^3}~|p_4|~
\left(\cos \theta ~A^+({\vf p};s)A({\vf p};s)+\sin \theta ~A^+(-{\vf
p};s)B^+({\vf p};s)+\right.\\
&&~~~~~~~~~~~~~~~~~~~~~~~~+\left.\sin \theta ~B(-{\vf p};s)A({\vf p};s)-
\cos \theta~ B({\vf p};s)B^+({\vf p};s)\right)~.\nonumber
\end{eqnarray}
There are two diagonal matrix elements of free Hamiltonian and now we are going
to consider one of them
$$\langle N|~B({\vf p};s)~B^+({\vf p};s)~| N\rangle \sim
\langle\sigma| A({\vf P};S)~B({\vf p};s)~B^+({\vf p};s)~A^+({\vf
P};S)|\sigma\rangle~.$$
While presenting the matrix element we demonstrate in the right-hand part only
one
partial contribution constructed by operator $A$ of some detached sort.
The others as it can easily be seen give a sum (integral) over states filling
the Fermi sphere in. Taking into account the normalization condition which we
assess as $\langle\sigma| A~A^+|\sigma\rangle=1$ (for operators with coinciding
arguments
$A$, $A^+$), the matrix element leads to the vacuum contribution similar to one
known from Ref. \cite{MZ}
$$- \int  \frat{d {\vf p}}{(2\pi)^3}~
\langle N|~|p_4|~\cos \theta~B({\vf p};s)~B^+({\vf p};s)~|N\rangle=
-\int \frat{d {\vf p}}{(2\pi)^3}~|p_4|~\cos\theta~.$$
We should also remember that $B |\sigma\rangle=0$, $A |\sigma\rangle=0$.
The partial contribution to the second matrix element equals to
\begin{equation}
\label{mtr2}
\langle\sigma| A({\vf P};S)~A^+({\vf p};s)~A({\vf p};s)~A^+({\vf
P};S)|\sigma\rangle
=(2\pi)^3~\delta({\vf p}-{\vf P})~\delta_{sS}~\langle\sigma| A({\vf
P};S)~A^+({\vf p};s)|\sigma\rangle~.
\end{equation}
Having filled  the state this contribution occurs as
$$ \int  \frat{d {\vf p}}{(2\pi)^3}~\langle N|~|p_4|~\cos \theta~A^+({\vf p};s)
~A({\vf p};s)~|N\rangle=\int^{P_F} \frat{d {\vf
p}}{(2\pi)^3}~|p_4|~\cos\theta~.$$

There are other diagonal matrix elements in the interaction Hamiltonian
$\bar q~t^a\gamma_\mu~q~\bar q'~t^b\gamma_\nu~q'$
which we are going to designate as 1), 2), $\alpha$),
$\beta$), $\gamma$), $\delta$). For example, 1) is the following matrix element
$\langle N| B~B^+~B'~B'^+|N\rangle$, 2) corresponds to $\langle N|
B~A~A'^+~B'^+| N\rangle$, $\alpha$) looks
like $\langle N| A~A^+~A'~A'^+|N\rangle$, $\beta$) is presented by $\langle N|
A~A^+~B'~B'^+|N\rangle$, $\gamma$) corresponds to $\langle N|
A^+~B^+~B'~A'|N\rangle$ and $\delta$) looks
like $\langle N| B~B'^+~A'^+~A'|N\rangle$.
The contribution $1)$ to the interaction term
$\langle N|~\bar q~t^a\gamma_\mu~q~\bar q'~t^b\gamma_\nu~q'~| N\rangle$
leads to the following spinor form
$$\overline{V}_{\alpha i}({\vf p},s)~t^a_{ij}\gamma^\mu_{\alpha \beta}
V_{\beta j}({\vf p},s)~\overline{V}_{\gamma k}({\vf p}',s')~t^b_{kl}
\gamma^\mu_{\gamma \delta}V_{\delta l}({\vf p}',s')~.$$
Bearing in mind the completeness property of spinor basis there appears the unit
matrix while summing up all colour polarization, for example,
$\sum_c V_i(c)\overline V_j(c)=\delta_{ij}$, and, therefore,
the contribution 1) as was mentioned in Ref. \cite{MZ}
turns out to be zero. The partial contribution $2)$ looks like
\begin{eqnarray}
\label{o4a}
&&\langle\sigma| A({\vf P};S)~B({\vf p};s)~A({\vf q};t)~A^+({\vf p}';s')
B^+({\vf q}';t')~A^+({\vf P};S)|\sigma\rangle=\nonumber\\
&&=(2\pi)^6\left[\delta({\vf q}-{\vf p}')\delta_{ts'}
\langle\sigma|A({\vf P};S)A^+({\vf P};S)|\sigma\rangle
-\delta({\vf q}-{\vf P})\delta_{tS}
\langle\sigma|A({\vf P};S)A^+({\vf p}';s')|\sigma\rangle
\right]\delta({\vf p}-{\vf q}')\delta_{st'}.\nonumber
\end{eqnarray}
One can obtain from this expression that it contributes to the matrix element
$\langle N|\bar q~t^a\gamma_\mu~q~\bar q'~t^b\gamma_\nu~q'~| N\rangle$
(if the normalization condition of one  single state is taken into account) with
the following result
\begin{eqnarray}
\label{m4a}
&&-\overline V_{\alpha i}({\vf p};s)~t^a_{ij}\gamma^\mu_{\alpha\beta}~U_{\beta
j}({\vf P};S)
\overline U_{\gamma k}({\vf P};S)~t^a_{kl}\gamma^\nu_{\gamma\delta}~V_{\delta
l}({\vf p};s)+\nonumber\\
&&
+\overline V_{\alpha i}({\vf p};s)~t^a_{ij}\gamma^\mu_{\alpha\beta}~U_{\beta
j}({\vf q};t)
\overline U_{\gamma k}({\vf q};t)~t^a_{kl}\gamma^\nu_{\gamma\delta}~V_{\delta
l}({\vf p};s)~.\nonumber
\end{eqnarray}
The calculation of matrix elements $\beta$), $\gamma$) and $\delta$)
demonstrates that they have a similar structure but their contributions cancel
each
other.

The matrix element $\alpha)$ deserves the special discussion.
Unlike the aforementioned examples the major partial contribution here
should be searched for the pair of quasiparticles
$$\langle\sigma|A({\vf Q};T) A({\vf P};S)~A^+({\vf p};s)~A({\vf q};t)~A^+({\vf
p}';s')
A({\vf q}';t')~A^+({\vf P};S)A^+({\vf Q};T)|\sigma\rangle~.$$
When the momenta ${\vf Q}$ and ${\vf P}$ coincide, as known, (see, for example,
Ref. \cite{adg}) the next term in the $1/V$ decomposition ($V$ is the volume
occupied by the system) appears. But if we are interested in knowing how one
quasiparticle modifies the dressing transformation it is necessary to consider
its matrix element with
$|1\rangle=A^+({\vf P};S)|\sigma\rangle$. Omitting the intermediate calculations
we present the contribution of the scattering term $\alpha)$ as
\begin{eqnarray}
\label{m4b}
&&-\overline U_{\alpha i}({\vf Q};T)~t^a_{ij}\gamma^\mu_{\alpha\beta}~U_{\beta
j}({\vf P};S)
\overline U_{\gamma k}({\vf P};S)~t^a_{kl}\gamma^\nu_{\gamma\delta}~U_{\delta
l}({\vf Q};T)+\nonumber\\
&&+\overline U_{\alpha i}({\vf P};S)~t^a_{ij}\gamma^\mu_{\alpha\beta}~U_{\beta
j}({\vf P};S)
\overline U_{\gamma k}({\vf Q};T)~t^a_{kl}\gamma^\nu_{\gamma\delta}~U_{\delta
l}({\vf Q};T)+\nonumber\\
&&+\overline U_{\alpha i}({\vf Q};T)~t^a_{ij}\gamma^\mu_{\alpha\beta}~U_{\beta
j}({\vf p};s)
\overline U_{\gamma k}({\vf p};s)~t^a_{kl}\gamma^\nu_{\gamma\delta}~U_{\delta
l}({\vf Q};T)+\nonumber\\
[-.2cm]
\\ [-.25cm]
&&+\overline U_{\alpha i}({\vf Q};T)~t^a_{ij}\gamma^\mu_{\alpha\beta}~U_{\beta
j}({\vf Q};T)
\overline U_{\gamma k}({\vf P};S)~t^a_{kl}\gamma^\nu_{\gamma\delta}~U_{\delta
l}({\vf P};S)+\nonumber\\
&&-\overline U_{\alpha i}({\vf P};S)~t^a_{ij}\gamma^\mu_{\alpha\beta}~U_{\beta
j}({\vf Q};T)
\overline U_{\gamma k}({\vf Q};T)~t^a_{kl}\gamma^\nu_{\gamma\delta}~U_{\delta
l}({\vf P};S)+\nonumber\\
&&+\overline U_{\alpha i}({\vf P};S)~t^a_{ij}\gamma^\mu_{\alpha\beta}~U_{\beta
j}({\vf p};s)
\overline U_{\gamma k}({\vf p};s)~t^a_{kl}\gamma^\nu_{\gamma\delta}~U_{\delta
l}({\vf P};S)~,\nonumber
\end{eqnarray}
The second and fourth terms give zero contribution for the same reasons
which were given for the contribution 1). The first and fifth terms, up to the
re-designations, are equal to each other. The third and sixth terms presents the
contribution of two states with momentum ${\vf P}$ and ${\vf Q}$. It is not
difficult to understand
that for the ensemble we get simply the integral over the Fermi sphere (see
below).

When one quasi-particle is considered the matrix element unlike (\ref{m4b}) is
proportional to
\begin{equation}
\label{m1}
\overline U_{\alpha i}({\vf P};S)~t^a_{ij}\gamma^\mu_{\alpha\beta}~U_{\beta
j}({\vf p};s)
\overline U_{\gamma k}({\vf p};s)~t^a_{kl}\gamma^\nu_{\gamma\delta}~U_{\delta
l}({\vf P};S)~.
\end{equation}
The polarization matrices of quasi-particles were considered in Ref. \cite{MZ}
and have the forms
\begin{eqnarray}
\label{22}
&&
V\overline{V}=p_4\gamma_4+\cos(\theta)~({\vf p}{\vf \gamma}-im)-
\frat{\stackrel{*}{\alpha}+\stackrel{}{\alpha}}{2im}~\sin(\theta)~
({\vf p}^2-im~{\vf p}{\vf \gamma})~,
\nonumber\\
&&U\overline{U}=p_4\gamma_4+\cos(\theta)~({\vf p}{\vf \gamma}+im)+
\frat{\stackrel{*}{\alpha}+\stackrel{}{\alpha}}{2im}~\sin(\theta)~
({\vf p}^2+im~{\vf p}{\vf \gamma})~,\nonumber
\end{eqnarray}
where the angle $\theta=2\varphi$, (the summation over polarization was done and
the matrix is diagonal in the colour indices). As a result for the traces which
we are interested in we receive
\begin{eqnarray}
\label{32}
&&\left.\begin{array}{l}
Tr~(V \overline V~ \gamma_\mu~t^a~U' \overline U' \gamma_\nu~t^a) \\
Tr~(U \overline U~ \gamma_\mu~t^a~U' \overline U' \gamma_\nu~t^a)\\
\end{array} \right\}
=4~\frat{N_c^2-1}{2}\times\nonumber\\ &&\times\left\{p_4 q_4~
g_{\mu\nu}\pm m^2~\delta_{\mu\nu} \left(\cos \theta -
\frat{\stackrel{*}{\alpha}+\stackrel{}{\alpha}}{2}~\frat{p^2}{m^2}\sin\theta\right)
\left(\cos \theta' -\frat{\stackrel{*}{\alpha}'
+\stackrel{}{\alpha}'}{2}~\frat{q^2}{m^2}\sin\theta'\right)+\right.\nonumber\\
[-.2cm]
\\ [-.25cm]
&&~~~+\left(\delta_{4\mu}\delta_{i\nu}+\delta_{4\nu}\delta_{i\mu}\right)
\left[\left(\cos \theta
+\frat{\stackrel{*}{\alpha}+\stackrel{}{\alpha}}{2}~\sin\theta\right)~q_4 p_i
+\left(\cos \theta'
+\frat{\stackrel{*}{\alpha}'+\stackrel{}{\alpha}'}{2}~\sin\theta'\right)~p_4
q_i\right]+\nonumber\\
&&~~~\left.+\left(\delta_{i\mu}\delta_{j\nu}-
\delta_{ij}\delta_{\mu\nu}+\delta_{i\nu}\delta_{j\mu}\right)~p_i q_j~
\left(\cos \theta
+\frat{\stackrel{*}{\alpha}+\stackrel{}{\alpha}}{2}~\sin\theta\right)
\left(\cos \theta' +\frat{\stackrel{*}{\alpha}'+\stackrel{}{\alpha}'}{2}
~\sin\theta'\right)\right\}~,\nonumber
\end{eqnarray}
where $p=|{\vf p}|$, $q=|{\vf q}|$ and $\theta'=\theta(q)$.
The coefficient $\alpha$ is defined up to the phase factor as
$\stackrel{*}{\alpha}\stackrel{}{\alpha}=m^2/{\vf p}^2$ which is quite
natural because a spinor has such a phase degree of freedom.
The analysis performed in Ref. \cite{MZ} demonstrates that the minimal value of
mean free energy is reached when this coefficient becomes the real number
$\alpha=\pm \frat{m}{|{\vf p}|}$.
For clarity we choose the positive sign and specify the form of correlation
function which is motivated by (anti-)instanton ensemble (for details see Ref.
\cite{MZ}),
\begin{equation}
\label{cor}
\langle A^{a}_\mu A^{b}_\nu\rangle({\vf x}-{\vf y})=\delta^{ab}~\widetilde G
~\frat{2}{N_c^2-1}~
\left[I({\vf x}-{\vf y})~\delta_{\mu\nu}-J_{\mu\nu}({\vf x}-{\vf y})\right]~.
\end{equation}
Here the second term is spanned onto the vector of relative distance.

Now we have to collect together all the results obtained for the one
quasi-particle. For the matrix element of interaction Hamiltonian we have
\begin{eqnarray}
\label{onep}
&&\langle 1|~\bar q~t^a~\gamma_\mu~q~\bar
q'~t^a~\gamma_\nu~q'~|1\rangle\sim\frat{N_c^2-1}{2}~
\frat{1}{4 |p_4| |q_4|}\times\nonumber\\[-.2cm]
\\ [-.25cm]
&&\times Tr\left[-V({\vf p})\overline V ({\vf p})\gamma_\mu U({\vf P})\overline
U ({\vf P})\gamma_\nu+
V({\vf p})\overline V ({\vf p})\gamma_\mu U({\vf q})\overline U ({\vf
q})\gamma_\nu +
U({\vf P})\overline U ({\vf P})\gamma_\mu U({\vf p})\overline U ({\vf
p})\gamma_\nu \right].\nonumber
\end{eqnarray}
The polarization indices are omitted here. It is interesting to notice that in
the first and third terms the factors spanned onto the tensors $g_{\mu\nu}$,
$(\delta_{4\mu}\delta_{i\nu}+\delta_{4\nu}\delta_{i\mu})$,
$(\delta_{i\mu}\delta_{j\nu}-
\delta_{ij}\delta_{\mu\nu}+\delta_{i\nu}\delta_{j\mu})$,
(see Eq. (\ref{32})) cancel each other. They survive only in the second term
which describes the pure vacuum contribution (considered in Ref. \cite{MZ}, see
also discussion below). Let us define the partial energy density per one quark
degree of freedom, as
$$w=\frat{{\cal E}}{2 N_c}~,~~~{\cal E}=E/V$$
where $E$ is the total energy of ensemble. Collecting all contributions together
we have for the state with the single quasi-particle
\begin{equation}
\label{onepe}
w_1=|p_4|~\cos\theta+2 G~\int \frat{d {\vf q}}{(2\pi)^3}~\frat{p q}{|p_4||q_4|}~
\left(\sin\theta-\frat{m}{p}\cos\theta\right)
\left(\sin\theta'-\frat{m}{q}\cos\theta'\right)~(I-J/4)+w_{vac}~,
\end{equation}
where $I=\widetilde I({\vf p}+{\vf q})$, $J_{ij}=J_{ij}({\vf p}+{\vf q})$,
$J=\sum_{i=1}^3J_{ii}$. The term $w_{vac}$ describes the vacuum contribution and
is shown below. It is convenient to pick out the colour index $G=\frat{2
\widetilde G}{N_c}$.
The energy of the state develops a minimal value if the condition
$$\frat{d w_1}{d\theta}=0~$$
is satisfied. Neglecting the modifications in the vacuum contribution
$w_{vac}$ we have
\begin{equation}
-|p_4|~\sin\theta+2 G~\frat{p}{|p_4|}~
\left(\sin\theta-\frat{m}{p}\cos\theta\right)
\int \frat{d {\vf q}}{(2\pi)^3}~\frat{q}{|q_4|}
\left(\sin\theta'-\frat{m}{q}\cos\theta'\right)~(I-J/4)=0~,
\end{equation}
which with the precision up to the terms $g_{\mu\nu}$,
$(\delta_{i\mu}\delta_{j\nu}-
\delta_{ij}\delta_{\mu\nu}+\delta_{i\nu}\delta_{j\mu})$,
is equivalent to the condition of minimum of mean vacuum energy
$\frat{d w_{vac}}{d\theta}=0$, see Ref. \cite{MZ}, i.e. in the situation of a
single quasi-particle the dressing transformation in this approximation
does not change compared to the vacuum one.
\begin{figure*}[!tbh]
\begin{center}
\includegraphics[width=0.5\textwidth]{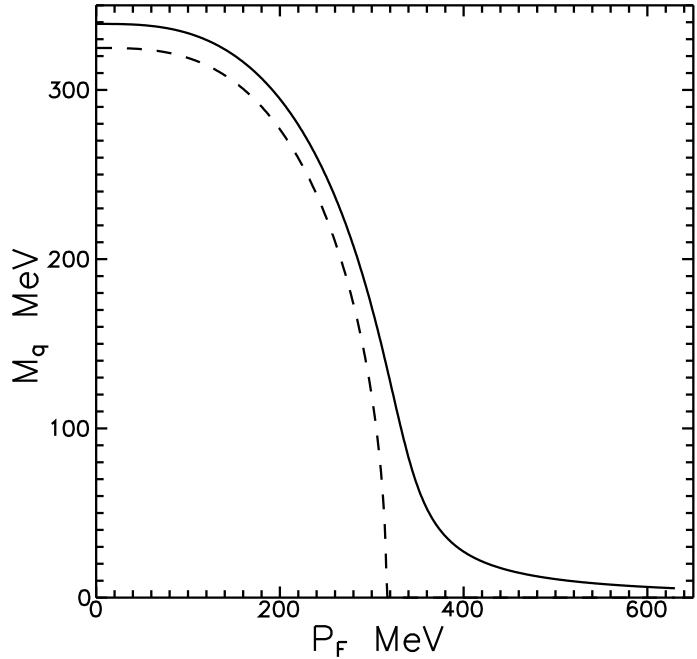}
\end{center}
\vspace{-7mm}
\caption{The dynamical quark mass ($|M_q|$) as a function of the Fermi momentum
for the NJL model. The solid line corresponds to the current quark mass
$m=5.5$ MeV. The dashed one shows the behaviour in the chiral limit.}
\label{f1}
\end{figure*}

For the matrix element of the interaction Hamiltonian for the ensemble of quasi-
particles we have
\begin{eqnarray}
\label{enp}
&&\langle N|~\bar q~t^a~\gamma_\mu~q~\bar q'~t^a~\gamma_\nu~q'~| N\rangle \sim
\frat{N_c^2-1}{2}~
\frat{1}{4 |p_4| |q_4|}\times\nonumber\\
&&\times Tr~\left[-V({\vf p})\overline V ({\vf p})~\gamma_\mu~U({\vf
P})\overline U ({\vf P})~\gamma_\nu+
V({\vf p})\overline V ({\vf p})~\gamma_\mu~U({\vf q})\overline U ({\vf
q})~\gamma_\nu +\right.\\
&&\left.+U({\vf P})\overline U ({\vf P})~\gamma_\mu~U({\vf p})\overline U ({\vf
p})~\gamma_\nu
+2~U({\vf P})\overline U ({\vf P})~\gamma_\mu~U({\vf Q})\overline U ({\vf
Q})~\gamma_\nu\right]~,\nonumber
\end{eqnarray}
(see Eq. (\ref{onep}) to compare). It is pertinent to mention the coefficient 2
in front of the last term. We present the matrix element implying some ordering
of the momenta, for
instance, $|{\vf Q}|<|{\vf P}|$. But in the expressions which are averaged over
the state $|N\rangle$ it is convenient to present the result by integrating over
the whole Fermi sphere, without taking into account the ordering. It is evident
then that we should take the contribution in factor two smaller in the
interaction
term, see the formula below.
In the last term similar to the 'vacuum' matrix element (the second term there)
the contributions spanned on the tensors  $g_{\mu\nu}$,
$(\delta_{i\mu}\delta_{j\nu}-
\delta_{ij}\delta_{\mu\nu}+\delta_{i\nu}\delta_{j\mu})$ survive.
For simplicity we consider in this paper only the situation when the second
correlator equals zero $J_{\mu\nu}=0$, besides we neglect all the
distinctions provoked by the tensors
$g_{\mu\nu}$, $(\delta_{i\mu}\delta_{j\nu}-
\delta_{ij}\delta_{\mu\nu}+\delta_{i\nu}\delta_{j\mu})$.
Collecting all the contributions together for the mean partial energy up to an
unessential constant we have
\begin{eqnarray}
\label{26}
&&\langle N|w| N\rangle=\int^{P_F}\frat{d {\vf p}}{(2\pi)^3}~ |p_4|
\cos\theta+\nonumber\\
&&+2 G\int^{P_F} \frat{d {\vf p}}{(2\pi)^3}~\frat{p}{|p_4|}~ \left(\sin\theta-
\frat{m}{p}\cos\theta\right)
\int \frat{d {\vf q}}{(2\pi)^3}~\frat{q}{|q_4|}~ \left(\sin\theta'-
\frat{m}{q}\cos\theta'\right)~I-\nonumber\\[-.2cm]
\\ [-.25cm]
&&-G\int^{P_F} \frat{d {\vf p}}{(2\pi)^3}~\frat{p}{|p_4|}~ \left(\sin\theta-
\frat{m}{p}\cos\theta\right)
\int^{P_F} \frat{d {\vf q}}{(2\pi)^3}~\frat{q}{|q_4|}~ \left(\sin\theta'-
\frat{m}{q}\cos\theta'\right)~I+\nonumber\\
&&+\int \frat{d {\vf p}}{(2\pi)^3}~|p_4|(1-\cos\theta)
-G\int \frat{d {\vf p}}{(2\pi)^3}~\frat{p}{|p_4|}\left(\sin\theta-
\frat{m}{p}\cos\theta\right)
\int \frat{d {\vf q}}{(2\pi)^3}~\frat{q}{|q_4|}\left(\sin\theta'-
\frat{m}{q}\cos\theta'\right) I~.\nonumber
\end{eqnarray}
To obtain this we perform a natural regularization ~subtracting the
contribution of free Hamiltonian  $H_0$. As a result there appears an unit
in the bracket containing  $-\cos\theta$ in the last line.
Performing the transposition of integration and changing the variables as ${\vf
p}$ $\to$ ${\vf q}$
the interaction term can be rewritten in the following conventional form
$2 \int^{P_F} \!\!d {\vf p}\int d {\vf q} -\int^{P_F}\!\!d {\vf p}\int^{P_F} d
{\vf q}-\int d {\vf p}\int d {\vf q}=-\int_{P_F}\!\! d {\vf p}\int_{P_F}\!\! d
{\vf q}$.
Finally we obtain the following expression for the partial energy
\begin{eqnarray}
\label{27}
&&\langle N|w| N\rangle=\int^{P_F}\frat{d {\vf p}}{(2\pi)^3}~ |p_4| +
\int_{P_F}\frat{d {\vf p}}{(2\pi)^3}~ |p_4|~(1-\cos\theta)-\nonumber\\[-.2cm]
\\ [-.25cm]
&&~~~~~~~~~~~~-G\int_{P_F} \frat{d {\vf p}}{(2\pi)^3}~\frat{p}{|p_4|}~
\left(\sin\theta-\frat{m}{p}\cos\theta\right)
\int_{P_F} \frat{d {\vf q}}{(2\pi)^3}~\frat{q}{|q_4|}~ \left(\sin\theta'-
\frat{m}{q}\cos\theta'\right)~I~.\nonumber
\end{eqnarray}
It has quite interesting interpretation. Compared to the vacuum mean
energy{\footnote{Just this expression is in the last line of Eq. (\ref{26}).}},
see. \cite{MZ},
\begin{eqnarray}
\label{evac}
&&w_{vac}=\langle \sigma|w|\sigma\rangle=\nonumber\\
&&=\int \frat{d {\vf p}}{(2\pi)^3}~|p_4|(1-\cos\theta)
-G\int \frat{d {\vf p}}{(2\pi)^3}~\frat{p}{|p_4|}\left(\sin\theta-
\frat{m}{p}\cos\theta\right)
\int \frat{d {\vf q}}{(2\pi)^3}~\frat{q}{|q_4|}\left(\sin\theta'-
\frat{m}{q}\cos\theta'\right)~I~.\nonumber
\end{eqnarray}
It is easy to see that in the considered symmetrical case when we neglect all
contributions generated by the tensors  $g_{\mu\nu}$,
$(\delta_{i\mu}\delta_{j\nu}-
\delta_{ij}\delta_{\mu\nu}+\delta_{i\nu}\delta_{j\mu})$,
for the state with the filled in Fermi sphere the angles of pairing could be
defined by the condition of functional minimum (\ref{27}) only for the momenta
larger than Fermi momentum $P_F$, see for comparison \cite{bern}.
The quarks forming the Fermi sphere look like the free (non-interacting) ones,
see the first term.
\begin{figure*}[!tbh]
\begin{center}
\includegraphics[width=0.5\textwidth]{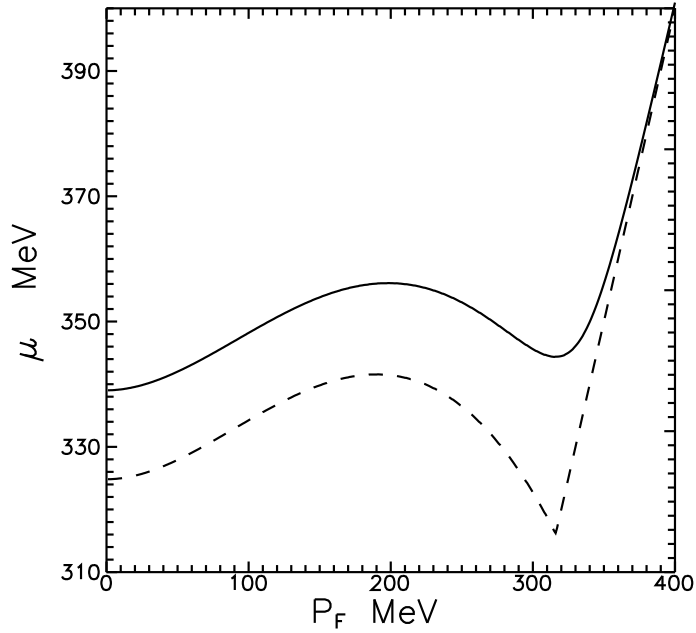}
\end{center}
\vspace{-7mm}
\caption{The quark chemical potential as a function of the Fermi momentum for
the NJL model. The solid line corresponds to the current quark mass $m=5.5$ MeV
and the dashed
one shows the behaviour in the chiral limit.}
\label{f2}
\end{figure*}
Now let us calculate the quark chemical potential which, by definition, is an
energy necessary for adding (removing) one quasi-particle to (from) a system
$\mu=\frat{\partial E}{\partial N}$, where
$N=2 N_c~ V~\int^{P_F}\frat{d {\vf p}}{(2\pi)^3}=\frat{N_c}{3\pi^2}~V~P_F^3$
is the total number of particles in the volume $V$.
Redefining the chemical potential as $\mu=\frat{2\pi^2}{P_F^2}~\frat{\partial
w}{\partial P_F}$.
we consider the model with correlation function behaving as the $\delta$-
function in the coordinate space. It is easy to see that we come to the Nambu--
Jona-
Lasinio
model (NJL) \cite{4} in this approach. The regularization is required to obtain
an
intelligent result in this model. We adjust the NJL model for the parameter set
given by Ref.\cite{5},
and limit the integration interval over momentum in Eq.(\ref{26}) with the
quantity $|{\vf p}|<\Lambda$ ($\Lambda=631$ MeV). Then the functional (\ref{27})
is written in the following form (inessential terms contributing  to the
constant
values are omitted)
\begin{equation}
\label{28}
w=w_0+\int^\Lambda_{P_F} \frat{d{\vf p}}{(2\pi)^3}~\left[|p_4|\left(1-
\cos\theta\right)
-G\frat{p}{|p_4|}\left(\sin\theta-\frat{m}{p}\cos\theta\right)\int^\Lambda_{P_F}
\frat{d{\vf q}}{(2\pi)^3}
\frat{q}{|q_4|}\left(\sin\theta'-\frat{m}{q}\cos\theta'\right)\right]~,
\end{equation}
where  $w_0=\int^{P_F} \frat{d{\vf p}}{(2\pi)^3}~|p_4|$ is the contribution
coming from the free quarks, $m=5.5$ MeV. The equation to calculate the
equilibrium angle $\theta$ reads as
\begin{equation}
\label{29}
(p^2+m^2)~\sin\theta-M_q\left(p\cos\theta+m\sin\theta\right)=0~,
\end{equation}
where
\begin{equation}
\label{30}
M_q=2G~\int^\Lambda_{P_F} \frat{d{\vf p}}{(2\pi)^3}\frat{p}{|p_4|}~
\left(\sin\theta-\frat{m}{p}\cos\theta\right)~.
\end{equation}
It allows us to obtain a well known selfconsistent gap equation for the
dynamical
quark mass.

For the parameters used the dynamical quark mass at zero Fermi momentum is
$M_q=-335$ MeV and for the quark condensate
\begin{equation}
\label{37}
\langle N|\bar q q| N\rangle=\frat{i~N_c}{\pi^2}~
\int^\Lambda_{P_F} dp~\frat{p^2}{|p_4|}~(p\sin\theta-m\cos\theta)~,
\end{equation}
we have $\langle\sigma|\bar q q|\sigma\rangle=-i~(247$ MeV)$^3$. The constant
characterizing a strength of four-fermion interaction was taken as
$G\Lambda^2/(2\pi^2)=1.34$. In Fig. \ref{f1} the dynamical quark mass as a
function of the Fermi
momentum is depicted. For comparison the data are presented for current quark
mass
$m=5.5$ MeV (the solid line) and the dashed line corresponds to the calculation
in the chiral limit. For the NJL model, in particular, the quark chemical
potential equals to
\begin{equation}
\label{munjl}
\mu=|P^F_4|~\cos\theta_F+M_q~\frat{P_F\sin\theta_F-m\cos\theta_F}{|P^F_4|}~.
\end{equation}
When the Fermi momentum reaches zero value the chemical potential quantity
coincides with dynamical quark mass $\mu(0)=|M_q|-M_q~m/|M_q|=|M_q|+m$.
In Fig. \ref{f2} the quark chemical potential is depicted as a function
of the Fermi momentum for the configurations analogous to the ones shown in Fig.
\ref{f1}. The chemical potential dependence on the Fermi momentum showing up
could be
interpreted as the effect of rapid decrease of the dynamical quark mass with the
Fermi
momentum increasing. Then using Eq.(\ref{29}) the chemical potential may be
presented
as the  following
$$\mu=\frat{M_q ~P_F}{|P^F_4| \sin \theta_F}~,
$$
and taking into account the identity
\begin{equation}
\label{iden}
(|P_4|^2-M_q m)^2+M_q^2 P^2=[P^2+(m-M_q)^2]~|P_4|^2~,
\end{equation}
we come to the noteworthy definition of the chemical potential
$$\mu=[P_F^2+(m-M_q)^2]^{1/2}~.$$
\begin{figure*}[!tbh]
\begin{center}
\includegraphics[width=0.5\textwidth]{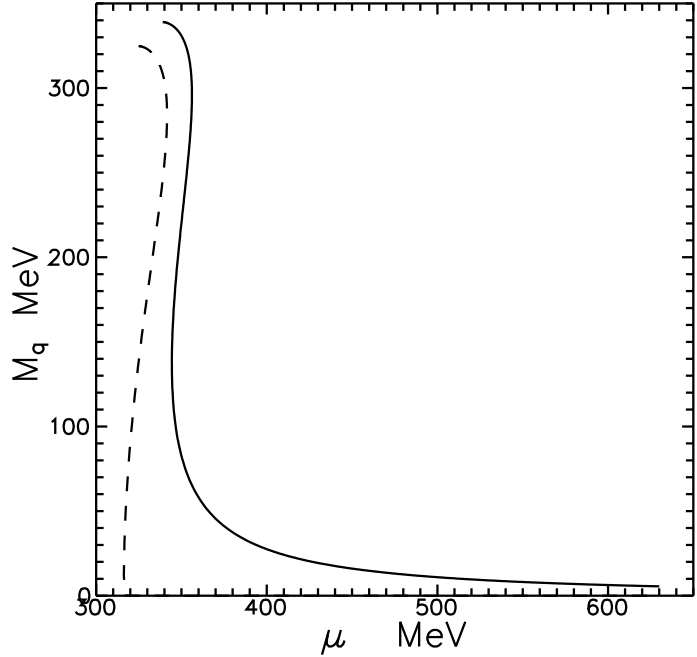}
\end{center}
\vspace{-7mm}
\caption{The dynamical quark mass ($|M_q|$) as a function of the chemical
potential. The solid line corresponds to the current quark mass $m=5.5$ MeV.
The dashed one shows the behaviour in the chiral limit.}
\label{f1b}
\end{figure*}
Let us remind that for the free fermion gas the chemical potential increases
monotonically with the Fermi momentum growing. The curious feature of the NJL
model is the appearance of state almost degenerate with vacuum state while the
process of filling up the Fermi sphere reaches to the momenta close to the
dynamical quark mass value (the similar value is characteristic for
the momentum of quark inside a baryon, see, for example, \cite{bub}).
This state density with the factor $3$ (which expresses the relation between
baryonic and quark degrees of freedom) absorbed corresponds to a normal nuclear
density
($n\sim 0.12$/fm$^3$), and chiral condensate could be estimated as $|<\bar q
q>^{1/3}|\sim 100$ MeV.
In the chiral limit the chemical potential near the discussed point is even
smaller
than the vacuum one. The full coincidence of the chemical potentials occurs at
the values of current quark mass around $2$ MeV. In fact, Fig. 2 shows that the
$u$ quark bond looks
stronger than one of the $d$ quark. For clarity in Fig. \ref{f1b} the dynamical
quark mass,
is plotted as the function of chemical potential (Fig. \ref{f1c} for the quark
condensate).
\begin{figure*}[!tbh]
\begin{center}
\includegraphics[width=0.5\textwidth]{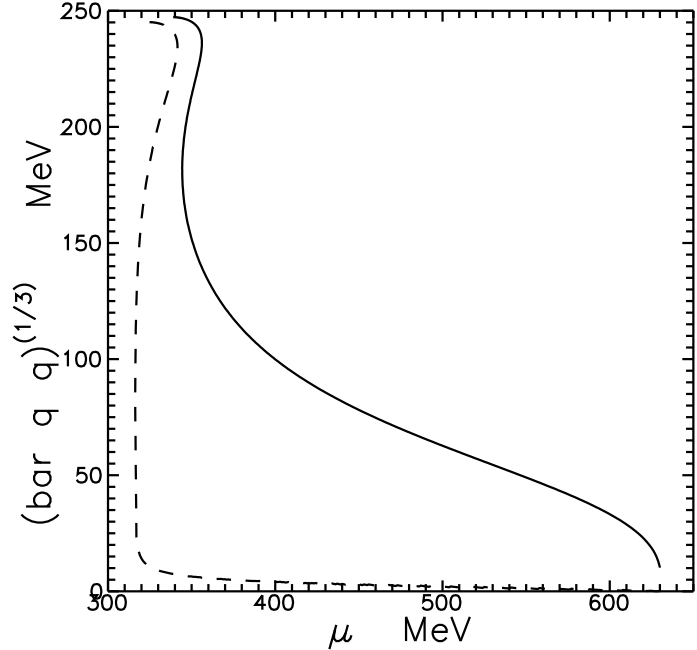}
\end{center}
\vspace{-7mm}
\caption{The quark condensate ($|<\bar q q>^{1/3}|$) as a function of the
chemical
potential. The solid line corresponds to the current quark mass $m=5.5$ MeV.
The dashed one shows the behaviour in the chiral limit.}
\label{f1c}
\end{figure*}

\begin{figure*}[!tbh]
\begin{center}
\includegraphics[width=0.5\textwidth]{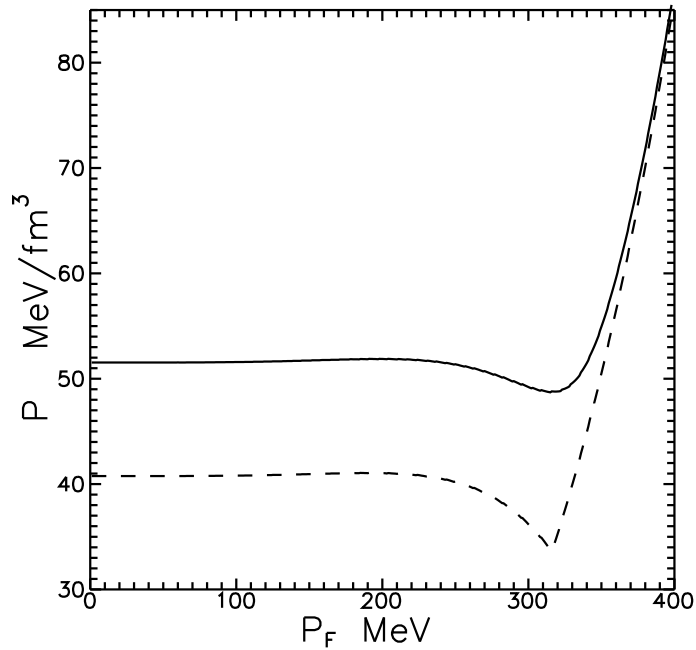}
\end{center}
\vspace{-7mm}
\caption{The pressure of the quark ensemble as function of the Fermi momentum.
The solid line for the current quark mass $m=5.5$ MeV. The dashed one---in
chiral limit.}
\label{f3}
\end{figure*}

The pressure of the quark ensemble
\begin{equation}
\label{press}
P=-\frat{d E}{d V}=-\frat{\partial E}{\partial V}+\frat{P_F}{3V}~\frat{\partial
E}{\partial P_F}=
 -{\cal E}+\mu~n~,
\end{equation}
is depicted on the Fig. \ref{f3} as a function of the Fermi momentum where
$n=N/V$ is the quark density. The quark pressure at the values of the Fermi
momentum close to the
quantity of dynamical quark mass is approximately degenerate with the vacuum
pressure
(slightly lower than the vacuum one). The vacuum density is of order  $40$---
$50$ MeV/fm$^3$ and
corresponds well to the value extracted from the bag models, see, for example,
Ref.
\cite{bub}.
Actually, all the NJL results could be received in the mean field approximation
because the trigonometrical terms in the mean energy definition  Eq.(\ref{28})
can be rewritten (using Eq.(\ref{iden}) again)  as the functions of dynamical
and current quark
masses in the following form
\begin{eqnarray}
\label{trig}
&&\frat{p~\sin\theta-m~\cos\theta}{|p_4|}=\frat{M_q-m}{[p^2+(M_q-
m)^2]^{1/2}}~,\nonumber\\
&&|p_4|~\cos\theta=\frat{p^2+m(m-M_q)}{[p^2+(M_q-m)^2]^{1/2}}~.\nonumber
\end{eqnarray}

In order to trace back the dependence of all results on the formfactor form we
consider the model (in a sense, opposite to the NJL model) with the formfactor
behaving
as $\delta$-function in momentum space $I(p)=(2 \pi)^3~\delta({\vf p})$. This
limit is an analogue of
the Keldysh model which is well known in the physics of condensed matter
\cite{6} and the mean
energy functional (\ref{27}) develops the following form
\begin{equation}
\label{k1}
w=\int^{P_F} \frat{d{\vf p}}{(2\pi)^3}~|p_4|
+\int_{P_F} \frat{d{\vf p}}{(2\pi)^3}~|p_4|\left(1-\cos\theta\right)
-G\int_{P_F} \frat{d{\vf p}}{(2\pi)^3}
\frat{p^2}{|p_4|^2}\left(\sin\theta-\frat{m}{p}\cos\theta\right)^2~.
\end{equation}
The chemical potential is defined in this approach as
\begin{equation}
\label{k2}
\mu=|P_4^F|~\cos\theta_F+G\frat{(P_F\sin\theta_F-m\cos\theta_F)^2}{|P_4^F|^2}~.
\end{equation}
It results from this definition that at low Fermi momenta the chemical potential
goes to $\mu \to m+G$. In Ref. \cite{MZ} the constant $G$ was taken of order of
dynamical quark mass in NJL. With Fermi momentum increasing the chemical
potential remains approximately constant and starts to increase when the Fermi
momentum exceeds
the $G$ value. The vacuum density ($P_F=0$) turns out to be singular and for the
pressure
difference we obtain
\begin{equation}
\label{k3}
P-P(0)=\frat{2~N_c}{2\pi^2}~
\left[\frat{P_F^3}{3}~\mu-
\int_0^{P_F}dp~ p^2 \cos\theta |p_4|-G
\int_0^{P_F}dp \frat{p^4}{|p_4|^2}\left(\sin\theta-
\frat{m}{p}\cos\theta\right)^2\right]~.
\end{equation}
This function is slowly monotonically growing till the Fermi momentum value of
$G$.

In summary we would like to emphasize our main result which, as we believe,is
quite
transparent. Our estimate of the effects responding to the process of filling up
the Fermi
sphere demonstrates the parallels between this quasi-particle picture
and the conventional bag model but with one new essential element. It is just
the
presence of instability region $dP/dP_F<0$. Then the states (\ref{7}) could be
considered
as natural 'building' material for baryons. In principle, one of the ways to
construct the
corresponding bound state could follow the Walecka model ideas \cite{wal} with
utilizing
the information on scalar and vector field behaviour and respective
constants of their interaction with quarks which have been gained up to now
\cite{bern}.

This work was supported by the INTAS Grant 04-84-398 and the Grant of National
Academy of Sciences of Ukraine.

\end{document}